\documentclass[a4paper,11pt]{article}
\usepackage{pos}
\usepackage{cases}
\usepackage[capitalize]{cleveref}

\newcommand{\C}[0]{\ensuremath{\mathbb{C}}}

\newcommand{\disk}[0]{\ensuremath{\mathbb{D}}}
\newcommand{\abs}[1]{\ensuremath{| #1|}}

\newtheorem{lemma}{Lemma}
\newtheorem{theorem}{Theorem}
\newtheorem{corollary}{Corollary}

\title{Approaching the Inverse Problem\\\normalsize{Toward Lattice QCD Calculations of Inclusive Hadronic Quantities}}
\ShortTitle{Approaching the Inverse Problem}

\author*[a,b]{William Jay}

\affiliation[a]{
Center for Theoretical Physics,
Massachusetts Institute of Technology, Cambridge, MA 02139, USA}

\affiliation[b]{Department of Physics, Colorado State University, Fort Collins, CO 80523, USA}

\emailAdd{william.jay@colostate.edu}

\DeclareMathOperator{\imag}{Im}

\abstract{
In this talk, I describe some recent ideas relating to the spectral reconstruction inverse problem, which arises frequently in lattice QCD calculations of inclusive hadronic quantities, and provide some physical context for this work.
Particular emphasis is given to a new method for rigorously bounding uncertainties using techniques from complex analysis.
}

\FullConference{
}

\begin{document}
\maketitle

\section{Motivation and context}

Numerical calculations using lattice quantum chromodynamics (QCD) provide correlation functions $G(\tau)$ at a discrete set of times Euclidean times $\tau$.
These correlation functions are related to spectral densities $\rho(\omega)$ via a Laplace transform, 
\begin{align} \label{eq:laplace}
G(\tau) = \int \frac{d\omega}{2\pi} \rho(\omega) e^{-\omega \tau}.
\end{align}
Computing the spectral density $\rho(\omega)$ given $G(\tau)$ amounts to an inverse Laplace transform, which is a subtle and difficult numerical problem.

Many important hadronic observables, including decay constants and form factors occurring in weak decays~\cite{FlavourLatticeAveragingGroupFLAG:2024oxs}, typically require knowledge of only the ground-state contribution to the spectral function.
In such problems, the challenge is reduced to clear isolation of the ground state.\footnote{
Even in these cases, robust quantification of excited-state effects can remain a difficult problem, especially for nucleon and certain other form factors.}
In contrast, the focus of my talk is on observables which really require some kind of knowledge about excited states. 
I will refer to this task of estimating $\rho(\omega)$ in \cref{eq:laplace} as \emph{the} inverse problem.

Phenomenological applications involving inclusive hadronic observables often require solution of the inverse problem---some knowledge of the full spectral function is needed because the hadronic states are ``summed over.''
Examples include the $R$-ratio for $e^+e^-\to{\rm hadrons}$~\cite{ExtendedTwistedMassCollaborationETMC:2022sta}, inclusive hadronic $\tau$ decay~\cite{ExtendedTwistedMass:2024myu,Evangelista:2023fmt}, inclusive semileptonic decays of $B$ mesons~\cite{Gambino:2022dvu,Kellermann:2024jqg,Barone:2023iat,Kellermann:2023yec,Barone:2023tbl,Gambino:2020crt}, inclusive neutrino-nucleus scattering~\cite{Fukaya:2020wpp,Liang:2023uai,Liang:2019frk}, and transport coefficients in hot QCD (a vast subject, reviewed recently in Ref~\cite{Rothkopf:2022fyo}).
Plenary talks at recent editions of this meeting have also dealt with topics closely related to the inverse problem, especially Refs.~\cite{Liang:2020sqi,Bulava:2023mjc}.


\section{Smeared spectral functions}

Intertwined conceptual and technical challenges render the inverse problem particularly difficult.
My talk highlighted two important aspects.
First, lattice QCD calculations in finite volume deform the spectrum.
Second, Euclidean data are available at a finite set of points.
Focusing on a smeared version of the problem, to be defined below, proves to be a fruitful approach to dealing with these difficulties.

Consider first the effect of a finite volume.
As for any quantum mechanical system ``in a box,'' the energy spectrum is discrete.
The situation contrasts sharply to physical scattering processes where a continuum of states appears above multi-particle thresholds.
This observation is the starting point for the finite-volume formalism, which relates finite-volume energy levels to infinite-volume scattering observables following seminal work in Ref.~\cite{Luscher:1990ck}.
For the present discussion, it suffices to say that the key inputs for these calculations are exact finite-volume energy levels.
Variational methods, in which one constructs an $n\times n$ matrix of correlation functions and then solves a generalized eigenvalue problem, are usually employed to extract $n$ such levels~\cite{Luscher:1990ux}, which then furnish the exact finite-volume spectral function through the $n$th level.
Instead, this talk focuses on smeared spectral functions, which provide an alternative and comparably new perspective on the connection to infinite-volume observables.

The second difficulty associated with the inverse problem arises from the fact that, although the values of $G(\tau)$ are known at perhaps one hundred Euclidean times, the spectral function $\rho(\omega)$ is desired along the real line.
One might image that a suitably smooth function function could, in fact, be well determined given such data.
However, spectral functions encountered in lattice QCD are not true functions in the mathematical sense but rather, being the sum of delta functions, are distributions.
A generic finite-volume spectral function has the form (for $\omega > 0$)
\begin{align}
    \rho_L(\omega) = \sum_{n=0}^\infty A_n \delta(\omega - E_n), \label{eq:finite-volume-spectral-function}
\end{align}
where $E_n$ and $A_n$ are the energy and amplitude of the $n$th state and $L$ denotes the linear extent of the finite spatial volume.
The standard method for approaching distributions is to consider integration against a suitable class of test functions.
A key insight in Ref.~\cite{Hansen:2017mnd} was to choose the test functions to be a family of smearing kernels $\hat{\delta}_\epsilon(\omega, \omega')$ satisfying 
\begin{align}
    \lim_{\epsilon\to 0} \hat{\delta}_\epsilon(\omega, \omega') = \delta(\omega - \omega'),
\end{align}
from which a smeared finite-volume spectral function is defined by convolution:
\begin{align}
    \hat{\rho}_L(\omega,\epsilon) = \int d\omega'\, 
    \hat{\delta}_\epsilon(\omega, \omega') \rho_L(\omega').
\end{align}
The infinite-volume spectral function can then be obtained as the result of an ordered limiting procedure:
\begin{align}
    \rho(\omega) = \lim_{\epsilon\to 0} \lim_{L\to\infty} \hat{\rho}_L(\omega, \epsilon).
\end{align}
Operationally, the right-hand side can be taken as a definition of the infinite-volume spectral function.
For the sake of concreteness, it is useful to keep a few particular smearing kernels in mind as examples.
The two that play an important role of the remainder of my talk are the Gaussian and Poisson (or Cauchy) kernels
\begin{subnumcases}
    {\hat{\delta}_\epsilon(\omega, \omega') =} 
        \frac{1}{\sqrt{2\pi}\epsilon} \exp\left( -\frac{(\omega-\omega')^2}{2\epsilon^2}\right), & {\rm Gaussian} \label{eq:gaussian}\\
        \frac{1}{\pi} \frac{\epsilon}{(\omega-\omega')^2+\epsilon^2}, & {\rm Poisson}. \label{eq:poisson}
\end{subnumcases}

Beside its mathematical utility, smearing has physical significance.
First of all, as recognized clearly in Ref.~\cite{Hansen:2017mnd}, a similar smearing kernel and ordered limit arises naturally in the standard derivation of Fermi's Golden rule; moreover, the precise form of the kernel is irrelevant in the calculation of total rates.
Second, the presence of the ordered limiting procedure resolves an otherwise confusing conceptual point:
On the one hand, for any finite volume, the spectral function consists of a discrete sum of delta functions.
On the other hand, for sufficiently large volumes, the delta functions should ``coalesce" above multi-particle thresholds to form a continuous line of singularities, i.e., a branch cut.
The ordered limiting procedure offers a concrete picture for how this ``coalescence'' happens.
By construction, the smeared spectral function is smooth for any volume.
Isolated delta functions on the real line from bound states are recovered as the smearing is removed by the second limit.

\section{The method of Hansen, Lupo, and Tantalo}

Prospects for lattice calculations of inclusive hadronic quantities received a boost in 2019, when Hansen, Lupo, and Tantalo (HLT)~\cite{Hansen:2019idp} rediscovered a variant of the Backus--Gilbert algorithm for spectral reconstruction~\cite{Backus:1968svk,Pijpers:1992}.
The method begins with a linear Ansatz for the smeared spectral function in terms of the input Euclidean data $G(\tau)$,
\begin{align}
    \hat{\rho}_L(\omega, \epsilon) = \sum_\tau g_\tau(\omega) G(\tau) = \int d\omega^\prime \rho(\omega^\prime) \hat{\delta}_\epsilon(\omega^\prime, \omega).
\end{align}
The unknown coefficients $g_\tau(\omega)$ are determined by minimizing the distance 
\begin{align}
A[q] = \int d\omega^\prime \left\{
        \delta_\epsilon(\omega^\prime - \omega) - 
        \hat{\delta}_\epsilon(\omega^\prime, \omega)
    \right\}^2
\end{align}
to some smearing kernel $\delta_\epsilon(\omega^\prime - \omega)$, which can be chosen freely.
The Gaussian kernel in \cref{eq:gaussian} is frequently a convenient choice.
In practice, the method actually minimizes a particular convex sum 
$\mathcal{F}_\lambda[q] = (1 - \lambda) A[g] + \lambda B[g]$,
where $B[q]$ is regulator term related to the covariance matrix of the input Euclidean data.
Roughly speaking, tuning the hyper-parameter $\lambda$ moves along a bias-variance trade-off curve.
An essential feature of the HLT method, both as described in Ref.~\cite{Hansen:2019idp} and refined later in Ref.~\cite{ExtendedTwistedMassCollaborationETMC:2022sta}, is the use of a stability analysis to look for a region where the reconstruction becomes insensitive to the choice of $\lambda$ and where errors are statistically dominated.
Ref.~\cite{LSDensities} provides an open-source implementation of the HLT algorithm in \texttt{python}.

An enlightening probabilistic interpretation of the HLT method using Bayesian inference and Gaussian processes was recently discussed in Refs.~\cite{Lupo:2023qna,DelDebbio:2024lwm}.
Gaussian-processes have also been used as a solution to the inverse problem independently elsewhere, e.g., Ref.~\cite{Pawlowski:2022zhh}.

As an illustrative example of the sorts of physical problems already being tackled using the HLT method, my talk highlighted a calculation of the smeared $R$-ratio for $e^+e^- \to {\rm hadrons}$ carried out by the Extended Twisted Mass Collaboration~\cite{ExtendedTwistedMassCollaborationETMC:2022sta}.
The calculation was completed using four 
gauge-field ensembles generated with four flavors of twisted-mass quarks with $a\approx 0.06-0.08$~fm and $M_\pi \approx 135$~MeV.
The authors applied the HLT algorithm to vector-current correlation functions using Gaussian smearing kernels with standard deviations of 0.44, 0.53, and 0.63~GeV to extract a smeared version of $R(s)$.
Two different volumes with $L\approx 5$ and $L\approx 7.5$~fm were employed at the coarsest lattice spacing
to give a explicit estimate of the finite-volume corrections, which was reported contribute insignificantly to the final statistical and systematic error budget.
The results were compared to the experimental results after applying the same finite smearing.
The theoretical results were sufficiently precise to report a roughly 3$\sigma$ tension with experiments for energies around the $\rho$ resonance.
Such a tension is interesting in light of recent theoretical~\cite{Aoyama:2020ynm} and experimental~\cite{Muong-2:2023cdq} work on the anomalous magnetic moment of the muon.
In the present context of my talk, the point I want to emphasize is that smeared observables, with \emph{finite} smearing, are often directly comparable to experimental data and can be of great phenomenological relevance.

The calculation in Ref.~\cite{ExtendedTwistedMassCollaborationETMC:2022sta} used rather broad smearing kernels.
An interesting question that arises is, as a matter of principle, ``How narrow could the smearing kernel be made while retaining systematic control of the reconstruction?'' 
In other words, for a given set of Euclidean data, ``How differential can one go?''
This question provides one motivation for the methods discussed next.

\section{Nevanlinna--Pick interpolation}

An alternative perspective on the inverse problem and the special role of smearing arises naturally in momentum space and was introduced in Ref.~\cite{Bergamaschi:2023xzx}.
It is well known, and reviewed accessibly in Ref.~\cite{Meyer:2011gj}, that the Fourier coefficients of $G(\tau)$ correspond to the momentum-space Green function evaluated at the Matsubara frequencies,
\begin{align}
    G(i \omega_l) = \int_0^\beta d\tau\, e^{i \omega_l \tau} G(\tau),
\end{align}
where $\omega_l = 2\ell\pi/\beta$ for bosons and $\omega_l = (2\ell+1)\pi/\beta$ for fermions.
In other words, the momentum-space Euclidean data amount to equally spaced points on the imaginary axis.
As usual, the spectral function is $\rho(\omega) = \frac{1}{\pi}\imag G(\omega)$.
The inverse problem now manifestly becomes one of analytic continuation: given values for the Green function on the imaginary axis, we seek to understand its behavior near the real line.
In Ref.~\cite{Bergamaschi:2023xzx}, it was shown that evaluation at any finite distance $\epsilon$ above the real line amounts to a smeared spectral function using the Poisson kernel of \cref{eq:poisson}, 
\begin{align}
\hat{\rho}_L(\omega, \epsilon) = \frac{1}{\pi} \imag G(\omega + i \epsilon).
\end{align}
In other words, analytic continuation is equivalent to computing a smeared spectral function using the Poisson kernel.
Essentially the same smearing was used with entirely different motivation in a classic paper by Poggio, Quinn, and Weinberg~\cite{Poggio:1975af}.

The problem of analytic continuation from a finite set of points does not typically have a unique solution, since many different analytic functions can interpolate a given set of points in the complex plane.
However, it turns out that methods from complex analysis nevertheless offer stringent and useful bounds to the task at hand.
The necessary results belong to Nevanlinna--Pick interpolation theory~\cite{Nevalinna1919,Nevalinna1929,Pick1915}, a century-old and highly developed branch of complex analysis~\cite{Nicolau2016,PickInterpolationBook,BlaschkeBook}.
Fortunately, a few elementary results will suffice for the present discussion.
The applicability of Nevanlinna--Pick interpolation to problems in field theory was first recognized in the context of condensed matter physics~\cite{PhysRevLett.126.056402}.

Recall that complex analytic functions are defined by convergent power series in an open set around each non-singular point.
For the momentum-space Green function $G(z)$, the singularities are confined to points $z=\pm E_n$ on the real line, so the power-series representation is possible at any point in the upper half plane and, in particular, in the neighborhood of the Euclidean data $G(i\omega_\ell)$.
However, the radius of convergence of the power series is determined by the distance to the nearest singularity, i.e., the distance to the ground state $E_0$.
The upshot is the familiar difficulty of ``seeing past the ground state,'' which asymptotically dominates a Euclidean correlation function.

This observation strongly suggests a change of coordinates on the domain $z=\omega+i\epsilon$ so that all points on the real line are roughly equidistant from the Euclidean data.
A well-known conformal map, the Cayley transform $C(z)$, does the trick: $C: \C^+ \to \disk$ with $C(z)=(z-i)/(z+i)$, where $\C^+$ denotes the upper half plane.
The Cayley transform takes the Euclidean data on the imaginary axis to the real line and the singularities on the real line to the boundary of the disk.

It is also advantageous to map the full codomain, and in particular the Euclidean data $G(i\omega_\ell)$, to the unit disk as well.
As first demonstrated in Ref.~\cite{PhysRevLett.126.056402}, for diagonal fermionic Green functions the Cayley transform suffices for this purpose.
For diagonal bosonic Green functions, a suitable conformal map is given in Ref.~\cite{Bergamaschi:2023xzx} (see also Ref.~\cite{Nogaki:2023mut} for a related treatment in the condensed-matter context).
The end result of these manipulations is that the Green function has been transformed into the natural object for study in complex analysis, namely, an analytic function from the disk to the disk.

The technical problem can now be specified precisely.
Given Euclidean data 
\begin{align}
\begin{split}
    \{i \omega_\ell\} &\mapsto \{\zeta_\ell\} \in \disk\\
    \{G(i\omega_\ell)\} &\mapsto \{w_\ell\} \in \disk,
\end{split}
\end{align}
construct an analytic function $f(\zeta):\disk \to \disk$ that interpolates these points, $f(\zeta_\ell) = w_\ell$.
Nevanlinna's Theorem solves this problem and will be given below.

The guiding credo of Nevanlinna--Pick interpolation can be stated easily, ``Factor out what you know.''
To implement this idea, we recall a useful property of analytic functions, which follows from the maximum modulus principle.
\begin{lemma}\label{thm:blaschke}
Let $g(\zeta): \disk \to \disk$ be an analytic function. Suppose that $g(\zeta)$ has a zero at the point $a\in \disk$, i.e., g(a)=0.
Then the original function can be written in factored form as $g(\zeta) = b_a(\zeta)\tilde{g}(\zeta)$, where $b_a(\zeta)$ is Blaschke factor which implements the zero and $\tilde{g}(\zeta)$ is a new analytic function.
\end{lemma}
Blaschke factors are familiar to our community, e.g., in the $z$-expansion of form factors appearing in quark-flavor physics, where they are used to factor out known analytic structure like sub-threshold poles~\cite{Boyd:1994tt,Boyd:1995cf,Boyd:1995sq,Boyd:1997kz,Grinstein:2017nlq,Caprini:1997mu}.
Repeated application of this factoring, which goes by the name of Schur's algorithm, leads to the theorem:
\begin{theorem}[Nevanlinna]\label{thm:nevanlinna}
Any solution to the interpolation problem with $N$ points, if it exists, can be written in the form
\begin{align}
    f(\zeta) = \frac{P_N(\zeta) f_N(\zeta) + Q_N(\zeta)}{R_N(\zeta) f_N(\zeta) + S_N(\zeta)},
\end{align}
where the functions $P_N$, $Q_N$, $R_N$, $S_N$ are known as the Nevanlinna coefficients and are calculable at any point $\zeta\in\disk$ from a known recursive formula in terms of the input data~\cite{Nicolau2016}.
The function $f_N(\zeta):\disk\to\disk$ is an arbitrary analytic function.
\end{theorem}
The appearance of the arbitrary function $f_N(\zeta)$ has a clear physical interpretation; it represents the freedom to specify additional input data to constrain further the interpolating function.
It plays the role of the ``remainder function'' $\tilde{g}(\zeta)$ in \cref{thm:blaschke}.
At first blush, the practical utility of \cref{thm:nevanlinna} is not obvious.
After all, what is to be done with the arbitrary function $f_N(\zeta)$?\footnote{
Ref.~\cite{PhysRevLett.126.056402}, which pioneered Nevanlinna--Pick interpolation for spectral reconstruction, has advocated treating the arbitrary function $f_N(\zeta)$ as a quantity to be optimized in order to impose a smoothness condition.
Absent further field-theoretic guidance, such an optimization seems unjustified and the likely source of uncontrolled systematic uncertainty.
}
However, the formulation of the problem on the unit disk pays another dividend.
The possible influence of the arbitrary function is necessarily constrained since $f_N(\zeta) \in \disk$.
The following corollary makes this observation precise.
\begin{corollary}[The Wertevorrat]\label{thm:wv}
For a given $\zeta\in\disk$, the space of all possible values that the interpolating function $f(\zeta)$ may take is given by a disk of radius $r_N(\zeta)$ centered at $c_N(\zeta)$, where
    \begin{align}
    c_N(\zeta)
    &=\frac
    {Q_N(\zeta) \overline{S}_N(\zeta) - P_N(\zeta) \overline{R}_N(\zeta)}
    {\abs{S_N(\zeta)}^2 - \abs{R_N(\zeta)}^2}
    \label{eq:wertevorrat_center}\\ 
    r_N(\zeta) 
    &= \frac{\abs{B_N(\zeta)}}{\abs{S_N(\zeta)}^2 - \abs{R_N(\zeta)}^2}. \label{eq:wertevorrat_boundary}
    \end{align}
    This disk is called the Wertevorrat $\Delta_N(\zeta)$.
\end{corollary}
The Wertevorrat is what gives \cref{thm:nevanlinna} practical utility for analytic continuation problems in lattice field theory.
Given $N$ interpolation points, the Wertevorrat $\Delta_N(\zeta)$ rigorously contains all possible analytic continuations at each extrapolation point $\zeta\in\disk$.
It offers a complete characterization of the systematic uncertainty associated with ``analytic continuation'' from a finite set of points.
Critically, the Wertevorrat is simply a consequence of the exactly known analytic structure of the problem and contains no model assumptions.
What's more, no ad hoc regularization is required (beyond the smearing).
Physical predictions follow after mapping the Wertevorrat from the disk back to the upper half-plane.
After transforming the result back to the upper half-plane, the total width of the imaginary part of the Wertevorrat gives the constraint on the smeared spectral function~\cite{Bergamaschi:2023xzx}.

Of course, invoking \cref{thm:nevanlinna} and \cref{thm:wv} depends on the existence of a solution in the first place.
Pick~\cite{Pick1915} has provided a necessary and sufficient condition for the existence of a solution.
\begin{theorem}[Pick]
    Given distinct $\{\zeta_n\}\in\disk$ and $\{w_n\}\in\disk$, there exists an analytic function $f:\disk\to\disk$ that interpolates the points ($f(\zeta_n)=w_n$) if and only if the Pick Matrix
    \begin{align}
        P_{ij} = \left[ \frac{1-w_i \bar{w}_j}{1-\zeta_i\bar{\zeta}_j}\right]_{1\leq i,j\leq n}
    \end{align}
    is positive semidefinite.
\end{theorem}
As observed in Ref.~\cite{PhysRevLett.126.056402} and reiterated in Ref.~\cite{Bergamaschi:2023xzx}, the presence of statistical uncertainties may cause this condition to fail.
Should this be the case, the construction of \cref{thm:nevanlinna} will still produce an interpolating function, but it will develop spurious poles somewhere in the disk.
To my knowledge, the question of how best to deal with this problem remains an interesting open question.
Important work in this direction includes Refs.~\cite{Huang:2022qsb,Yu:2024ncm}.
Of course, invoking the theorem without due care results in not-unexpected numerical instability~\cite{Huang:2023gpb}.

The Wertevorrat has several important properties.
First, when evaluated on the boundary of the disk ($\epsilon=0$), corresponding to the unsmeared spectral function, the Wertevorrat generically fills the full unit disk.
In the upper-half plane, this behavior corresponds to a complete lack of knowledge of the finite-volume spectral function and is the manifestation of the ``ill-posed'' nature of the inverse Laplace transform in the present formalism.
Second, when the Pick matrix becomes exactly singular ($\det P = 0$), the interpolation problem is said to be \emph{extremal}~\cite{Nicolau2016}.
For extremal problems, the Wertevorrat shrinks to a point for any $\zeta\in\disk$, which corresponds to vanishing uncertainty in the analytic continuation.
Physical examples of extremal problems include analytic continuation of a spectral decomposition truncated to $n$ states; 
one can show that such problems become extremal after interpolating $2n$ points.\footnote{This appealing property agrees with intuition: $2n$ points should, and in fact \emph{do}, provide enough information to determine completely the amplitudes and energies of $n$ states.}
Third, for a fixed set of input data, the Wertevorrat is observed empirically to grow roughly exponentially with the approach ($\epsilon \to 0$) to the boundary of the disk.
Finally, for a fixed extrapolation point $\zeta\in\disk$, the Wertevorrat is observed empirically to decrease roughly exponentially as the number of interpolation points increases.

\begin{figure}[t!]
    \centering
    \includegraphics[width=0.95\linewidth]{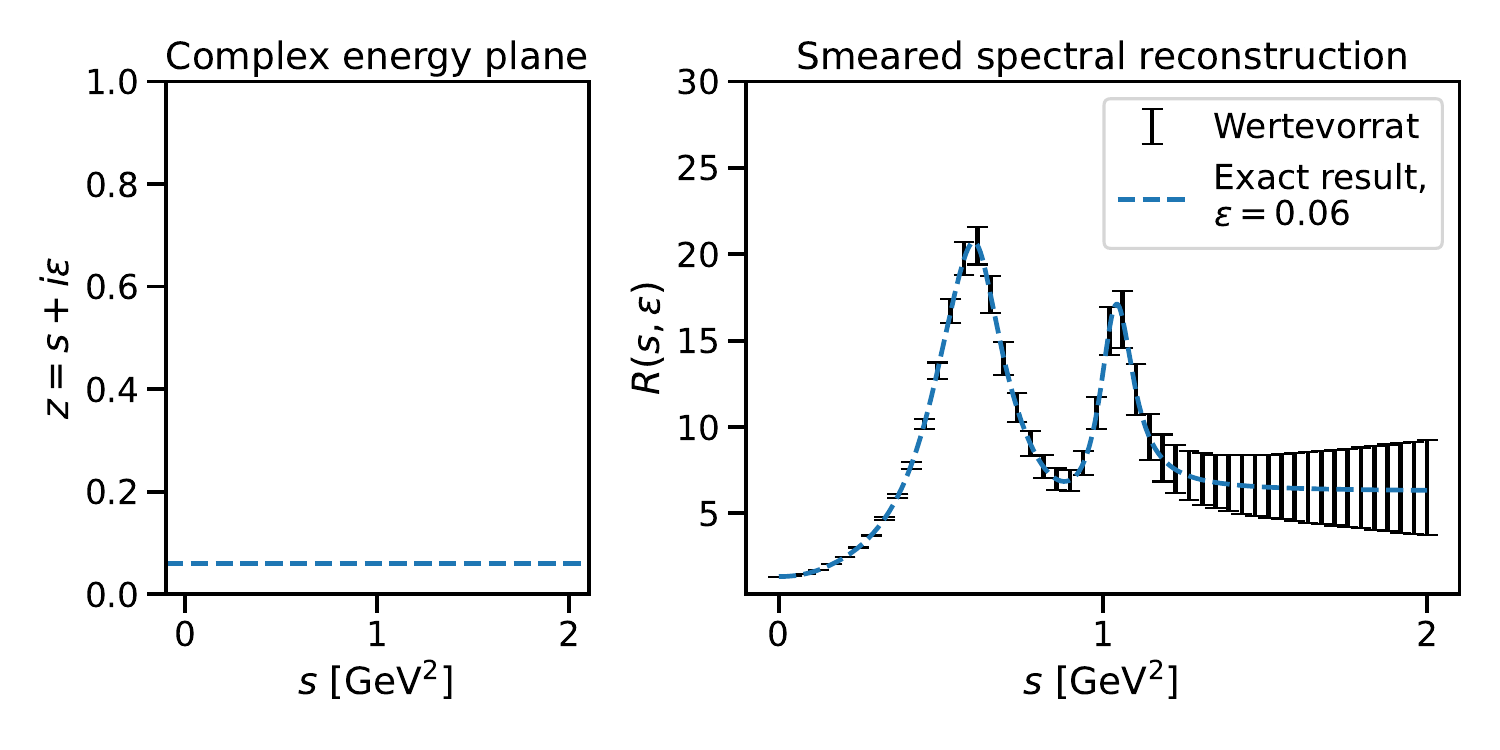}
    \caption{The reconstruction of the smeared $R$-ratio $R(s, \epsilon)$ using exact Euclidean data at $\beta=96$ total points generated according to the Bernecker--Meyer model for $R(s)$~\cite{Bernecker:2011gh}.
    The left panel shows the line $z=\omega + i 0.06$ in the complex plane upon which the smeared spectral function is evaluated.
    The right panel shows the known exact result for $R(s,\epsilon)$ (blue curve) as well as the result from the Wertevorrat (black points).
    }
    \label{fig:smeared_R-ratio}
\end{figure}

\begin{figure}[t!]
    \centering
    \includegraphics[width=0.65\linewidth]{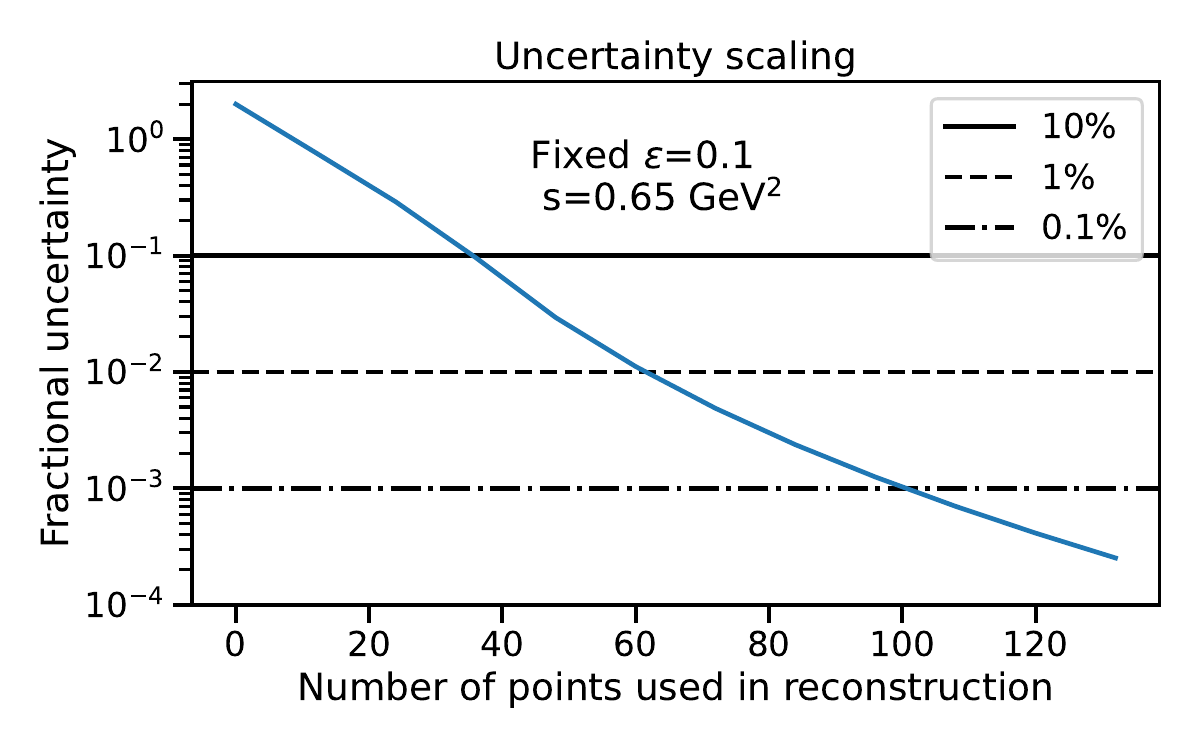}
    \caption{The scaling of the uncertainty in the smeared spectral reconstruction for $R(s, \epsilon)$ as the number of points in the interpolation is varied.
    The fractional uncertainty coming from the Wertevorrat decreases roughly exponentially.
    The result is for a fixed $z=\omega+i\epsilon$ near the $\rho$-meson peak in $R(s)$.
    For a smearing of $\epsilon=0.1$ at $s=0.65~{\rm GeV}^2$, 60 and 100 points suffice to determine $R(s,\epsilon)$ with percent and per mille precision, respectively.
    }
    \label{fig:scaling}
\end{figure}

As with the HLT method discussed above, the $R$-ratio also provides an interesting test system for the Wertevorrat.
The $R$-ratio was considered in some detail in Ref.~\cite{Bergamaschi:2023xzx} using the Bernecker--Meyer parameterization for $R(s)$~\cite{Bernecker:2011gh}.
An example of a smeared reconstruction for the $R$-ratio is given in \cref{fig:smeared_R-ratio}, where prominent peaks from the $\rho$ and $\phi$ resonances are clearly visible with modest uncertainty.
A new property discussed for the first time in this talk was the scaling of the Wertevorrat with the number of interpolation points.
\Cref{fig:scaling} shows the roughly exponential decrease in the Wertevorrat near the $\rho$ peak as the number of interpolating points increases.

\section{Conclusions}

In my talk, I described the phenomenological importance of inclusive hadronic quantities and the use of smeared spectral functions to access them.
A particular emphasis was given to the Wertevorrat arising in Nevanlinna--Pick interpolation, which offers a systematically improvable approach for increasing energy resolution in spectral reconstructions.
Crucially, the Wertevorrat bounds the full systematic uncertainty associated with ``analytic continuation'' from a finite set of points, even when this uncertainty is not small.

Unfortunately, my talk did not have time to touch on other exciting formal developments in spectral reconstruction, including Haag--Ruelle scattering theory~\cite{Patella:2024cto}, Mellin transforms~\cite{Bruno:2024fqc}, and Lanczos methods~\cite{Wagman:2024rid,Hackett:2024nbe,Hackett:2024xnx}.
My hope and expectation for the coming years is that the these ideas will be integrated alongside established methods to enable exciting, new calculations in QCD and beyond.

\acknowledgments
I am grateful to the members of the MIT lattice group for encouragement and valuable feedback during the preparation of this talk.
My understanding of the inverse problem has been deepened through conversations with many people; 
special thanks goes to
Ryan Abbott,
Dan Hackett,
Max Hansen,
Patrick Oare,
Mike Wagman,
Fernando Romero-L{\'o}pez, and
Julian Urban.
This work was supported in part by the US DOE Office of Science under grant Contract Numbers DE-SC0011090 and DE-SC0021006
and by the Simons Foundation grant 994314 (Simons Collaboration on Confinement and QCD Strings).

\bibliographystyle{JHEP}
\bibliography{refs}

\providecommand{\href}[2]{#2}\begingroup\raggedright\begin{thebibliography}{10}

\bibitem{FlavourLatticeAveragingGroupFLAG:2024oxs}
{\scshape Flavour Lattice Averaging Group (FLAG)} collaboration, \emph{{FLAG Review 2024}},  \href{https://arxiv.org/abs/2411.04268}{{\ttfamily 2411.04268}}.

\bibitem{ExtendedTwistedMassCollaborationETMC:2022sta}
{\scshape Extended Twisted Mass Collaboration (ETMC)} collaboration, \emph{{Probing the Energy-Smeared R Ratio Using Lattice QCD}}, \href{https://doi.org/10.1103/PhysRevLett.130.241901}{\emph{Phys. Rev. Lett.} {\bfseries 130} (2023) 241901} [\href{https://arxiv.org/abs/2212.08467}{{\ttfamily 2212.08467}}].

\bibitem{ExtendedTwistedMass:2024myu}
{\scshape Extended Twisted Mass} collaboration, \emph{{Inclusive Hadronic Decay Rate of the \ensuremath{\tau} Lepton from Lattice QCD: The \ensuremath{\bar{u}s} Flavor Channel and the Cabibbo Angle}}, \href{https://doi.org/10.1103/PhysRevLett.132.261901}{\emph{Phys. Rev. Lett.} {\bfseries 132} (2024) 261901} [\href{https://arxiv.org/abs/2403.05404}{{\ttfamily 2403.05404}}].

\bibitem{Evangelista:2023fmt}
{\scshape Extended Twisted Mass} collaboration, \emph{{Inclusive hadronic decay rate of the \ensuremath{\tau} lepton from lattice QCD}}, \href{https://doi.org/10.1103/PhysRevD.108.074513}{\emph{Phys. Rev. D} {\bfseries 108} (2023) 074513} [\href{https://arxiv.org/abs/2308.03125}{{\ttfamily 2308.03125}}].

\bibitem{Gambino:2022dvu}
P.~Gambino, S.~Hashimoto, S.~M\"achler, M.~Panero, F.~Sanfilippo, S.~Simula et~al., \emph{{Lattice QCD study of inclusive semileptonic decays of heavy mesons}}, \href{https://doi.org/10.1007/JHEP07(2022)083}{\emph{JHEP} {\bfseries 07} (2022) 083} [\href{https://arxiv.org/abs/2203.11762}{{\ttfamily 2203.11762}}].

\bibitem{Kellermann:2024jqg}
R.~Kellermann, A.~Barone, A.~Elgaziari, S.~Hashimoto, Z.~Hu, A.~J\"uttnerc et~al., \emph{{Systematic effects in the lattice calculation of inclusive semileptonic decays}},  in \emph{{41st International Symposium on Lattice Field Theory}}, 11, 2024 [\href{https://arxiv.org/abs/2411.18058}{{\ttfamily 2411.18058}}].

\bibitem{Barone:2023iat}
A.~Barone, S.~Hashimoto, A.~J\"uttner, T.~Kaneko and R.~Kellermann, \emph{{Chebyshev and Backus-Gilbert reconstruction for inclusive semileptonic $B_{(s)}$-meson decays from Lattice QCD}}, \href{https://doi.org/10.22323/1.453.0236}{\emph{PoS} {\bfseries LATTICE2023} (2024) 236} [\href{https://arxiv.org/abs/2312.17401}{{\ttfamily 2312.17401}}].

\bibitem{Kellermann:2023yec}
R.~Kellermann, A.~Barone, S.~Hashimoto, A.~J\"uttner\ensuremath{\mathit{c}} and T.~Kaneko\ensuremath{\mathit{a}}, \emph{{Studies on finite-volume effects in the inclusive semileptonic decays of charmed mesons}}, \href{https://doi.org/10.22323/1.453.0272}{\emph{PoS} {\bfseries LATTICE2023} (2024) 272} [\href{https://arxiv.org/abs/2312.16442}{{\ttfamily 2312.16442}}].

\bibitem{Barone:2023tbl}
A.~Barone, S.~Hashimoto, A.~J\"uttner, T.~Kaneko and R.~Kellermann, \emph{{Approaches to inclusive semileptonic B$_{(s)}$-meson decays from Lattice QCD}}, \href{https://doi.org/10.1007/JHEP07(2023)145}{\emph{JHEP} {\bfseries 07} (2023) 145} [\href{https://arxiv.org/abs/2305.14092}{{\ttfamily 2305.14092}}].

\bibitem{Gambino:2020crt}
P.~Gambino and S.~Hashimoto, \emph{{Inclusive Semileptonic Decays from Lattice QCD}}, \href{https://doi.org/10.1103/PhysRevLett.125.032001}{\emph{Phys. Rev. Lett.} {\bfseries 125} (2020) 032001} [\href{https://arxiv.org/abs/2005.13730}{{\ttfamily 2005.13730}}].

\bibitem{Fukaya:2020wpp}
H.~Fukaya, S.~Hashimoto, T.~Kaneko and H.~Ohki, \emph{{Towards fully nonperturbative computations of inelastic $\ell N$ scattering cross sections from lattice QCD}}, \href{https://doi.org/10.1103/PhysRevD.102.114516}{\emph{Phys. Rev. D} {\bfseries 102} (2020) 114516} [\href{https://arxiv.org/abs/2010.01253}{{\ttfamily 2010.01253}}].

\bibitem{Liang:2023uai}
J.~Liang, R.S.~Sufian, B.~Wang, T.~Draper, T.~Khan, K.-F.~Liu et~al., \emph{{Elastic and resonance structures of the nucleon from hadronic tensor in lattice QCD: implications for neutrino-nucleon scattering and hadron physics}},  \href{https://arxiv.org/abs/2311.04206}{{\ttfamily 2311.04206}}.

\bibitem{Liang:2019frk}
{\scshape XQCD} collaboration, \emph{{Towards the nucleon hadronic tensor from lattice QCD}}, \href{https://doi.org/10.1103/PhysRevD.101.114503}{\emph{Phys. Rev. D} {\bfseries 101} (2020) 114503} [\href{https://arxiv.org/abs/1906.05312}{{\ttfamily 1906.05312}}].

\bibitem{Rothkopf:2022fyo}
A.~Rothkopf, \emph{{Inverse problems, real-time dynamics and lattice simulations}}, \href{https://doi.org/10.1051/epjconf/202227401004}{\emph{EPJ Web Conf.} {\bfseries 274} (2022) 01004} [\href{https://arxiv.org/abs/2211.10680}{{\ttfamily 2211.10680}}].

\bibitem{Liang:2020sqi}
{\scshape \ensuremath{\chi}QCD} collaboration, \emph{{PDFs and Neutrino-Nucleon Scattering from Hadronic Tensor}}, \href{https://doi.org/10.22323/1.363.0046}{\emph{PoS} {\bfseries LATTICE2019} (2020) 046} [\href{https://arxiv.org/abs/2008.12389}{{\ttfamily 2008.12389}}].

\bibitem{Bulava:2023mjc}
J.~Bulava, \emph{{The spectral reconstruction of inclusive rates}}, \href{https://doi.org/10.22323/1.430.0231}{\emph{PoS} {\bfseries LATTICE2022} (2023) 231} [\href{https://arxiv.org/abs/2301.04072}{{\ttfamily 2301.04072}}].

\bibitem{Luscher:1990ck}
M.~Luscher and U.~Wolff, \emph{{How to Calculate the Elastic Scattering Matrix in Two-dimensional Quantum Field Theories by Numerical Simulation}}, \href{https://doi.org/10.1016/0550-3213(90)90540-T}{\emph{Nucl. Phys. B} {\bfseries 339} (1990) 222}.

\bibitem{Luscher:1990ux}
M.~Luscher, \emph{{Two particle states on a torus and their relation to the scattering matrix}}, \href{https://doi.org/10.1016/0550-3213(91)90366-6}{\emph{Nucl. Phys. B} {\bfseries 354} (1991) 531}.

\bibitem{Hansen:2017mnd}
M.T.~Hansen, H.B.~Meyer and D.~Robaina, \emph{{From deep inelastic scattering to heavy-flavor semileptonic decays: Total rates into multihadron final states from lattice QCD}}, \href{https://doi.org/10.1103/PhysRevD.96.094513}{\emph{Phys. Rev. D} {\bfseries 96} (2017) 094513} [\href{https://arxiv.org/abs/1704.08993}{{\ttfamily 1704.08993}}].

\bibitem{Hansen:2019idp}
M.~Hansen, A.~Lupo and N.~Tantalo, \emph{{Extraction of spectral densities from lattice correlators}}, \href{https://doi.org/10.1103/PhysRevD.99.094508}{\emph{Phys. Rev. D} {\bfseries 99} (2019) 094508} [\href{https://arxiv.org/abs/1903.06476}{{\ttfamily 1903.06476}}].

\bibitem{Backus:1968svk}
G.~Backus and F.~Gilbert, \emph{{The Resolving Power of Gross Earth Data}}, \href{https://doi.org/10.1111/j.1365-246X.1968.tb00216.x}{\emph{Geophys. J. Int.} {\bfseries 16} (1968) 169}.

\bibitem{Pijpers:1992}
F.~Pijpers and M.~Thompson, \emph{{Faster formulations of the optimally localized averages method for helioseismic inversions}}, \href{https://doi.org/10.1103/RevModPhys.87.1067}{\emph{Astronomy and Astrophysics} {\bfseries 262} (1992) L33}.

\bibitem{LSDensities}
A.~Lupo and N.~Forzano, ``Lsdensities: Lattice spectral densities.'' \url{https://github.com/LupoA/lsdensities}.

\bibitem{Lupo:2023qna}
A.~Lupo, L.~Del~Debbio, M.~Panero and N.~Tantalo, \emph{{Bayesian interpretation of Backus-Gilbert methods}}, \href{https://doi.org/10.22323/1.453.0004}{\emph{PoS} {\bfseries LATTICE2023} (2024) 004} [\href{https://arxiv.org/abs/2311.18125}{{\ttfamily 2311.18125}}].

\bibitem{DelDebbio:2024lwm}
L.~Del~Debbio, A.~Lupo, M.~Panero and N.~Tantalo, \emph{{Bayesian solution to the inverse problem and its relation to Backus-Gilbert methods}},  \href{https://arxiv.org/abs/2409.04413}{{\ttfamily 2409.04413}}.

\bibitem{Pawlowski:2022zhh}
J.M.~Pawlowski, C.S.~Schneider, J.~Turnwald, J.M.~Urban and N.~Wink, \emph{{Yang-Mills glueball masses from spectral reconstruction}}, \href{https://doi.org/10.1103/PhysRevD.108.076018}{\emph{Phys. Rev. D} {\bfseries 108} (2023) 076018} [\href{https://arxiv.org/abs/2212.01113}{{\ttfamily 2212.01113}}].

\bibitem{Aoyama:2020ynm}
T.~Aoyama et~al., \emph{{The anomalous magnetic moment of the muon in the Standard Model}}, \href{https://doi.org/10.1016/j.physrep.2020.07.006}{\emph{Phys. Rept.} {\bfseries 887} (2020) 1} [\href{https://arxiv.org/abs/2006.04822}{{\ttfamily 2006.04822}}].

\bibitem{Muong-2:2023cdq}
{\scshape Muon g-2} collaboration, \emph{{Measurement of the Positive Muon Anomalous Magnetic Moment to 0.20~ppm}}, \href{https://doi.org/10.1103/PhysRevLett.131.161802}{\emph{Phys. Rev. Lett.} {\bfseries 131} (2023) 161802} [\href{https://arxiv.org/abs/2308.06230}{{\ttfamily 2308.06230}}].

\bibitem{Bergamaschi:2023xzx}
T.~Bergamaschi, W.I.~Jay and P.R.~Oare, \emph{{Hadronic structure, conformal maps, and analytic continuation}}, \href{https://doi.org/10.1103/PhysRevD.108.074516}{\emph{Phys. Rev. D} {\bfseries 108} (2023) 074516} [\href{https://arxiv.org/abs/2305.16190}{{\ttfamily 2305.16190}}].

\bibitem{Meyer:2011gj}
H.B.~Meyer, \emph{{Transport Properties of the Quark-Gluon Plasma: A Lattice QCD Perspective}}, \href{https://doi.org/10.1140/epja/i2011-11086-3}{\emph{Eur. Phys. J. A} {\bfseries 47} (2011) 86} [\href{https://arxiv.org/abs/1104.3708}{{\ttfamily 1104.3708}}].

\bibitem{Poggio:1975af}
E.C.~Poggio, H.R.~Quinn and S.~Weinberg, \emph{{Smearing the Quark Model}}, \href{https://doi.org/10.1103/PhysRevD.13.1958}{\emph{Phys. Rev. D} {\bfseries 13} (1976) 1958}.

\bibitem{Nevalinna1919}
R.~Nevanlinna, \emph{{{\"U}ber beschr{\"a}nkte Funktionen die in gegebenen punkten vorgeschriebene Werte annehmen}}, {\emph{Ann. Acad. Sci. Fenn. Ser. A} {\bfseries 13} (1919) }.

\bibitem{Nevalinna1929}
R.~Nevanlinna, \emph{{{\"U}ber beschr{\"a}nkte analytische Funktionen}}, {\emph{Ann. Acad. Sci. Fenn. Ser. A} {\bfseries 32} (1929) }.

\bibitem{Pick1915}
G.~Pick, \emph{{{\"U}ber die Beschr{\"a}nkungen analytischer Funktionen, welche durch vorgegebene Funktionswerte bewirkt werden}}, \href{https://doi.org/10.1007/BF01456817}{\emph{Math. Ann.} {\bfseries 77} (1915) 7}.

\bibitem{Nicolau2016}
A.~Nicolau, \emph{{The Nevanlinna-Pick Interpolation Problem}},  in \emph{Proceedings of the Summer School in Complex and Harmonic analysis, and related topics}, J.~Gr{\"o}hn, J.~Heittokangas, R.~Korhonen and J.~R{\"a}tty{\"a}, eds., no.~22 in Reports and Studies in Forestry and Natural Sciences, Publications of the University of eastern Finland, 2016, \href{https://erepo.uef.fi/handle/123456789/15782}{https://erepo.uef.fi/handle/123456789/15782}.

\bibitem{PickInterpolationBook}
J.~Agler and J.~McCarthy, \emph{Pick Interpolation and Hilbert Function Spaces}, American Mathematical Society (2002), \href{https://doi.org/10.1090/gsm/044}{10.1090/gsm/044}.

\bibitem{BlaschkeBook}
S.~Garcia, J.~Mashreghi and W.~Ross, \emph{Finite Blashke Products and Their Connections}, Springer, 1~ed. (May, 2018), \href{https://doi.org/10.1007/978-3-319-78247-8}{10.1007/978-3-319-78247-8}.

\bibitem{PhysRevLett.126.056402}
J.~Fei, C.-N.~Yeh and E.~Gull, \emph{Nevanlinna analytical continuation}, \href{https://doi.org/10.1103/PhysRevLett.126.056402}{\emph{Phys. Rev. Lett.} {\bfseries 126} (2021) 056402}.

\bibitem{Nogaki:2023mut}
K.~Nogaki and H.~Shinaoka, \emph{{Bosonic Nevanlinna Analytic Continuation}}, \href{https://doi.org/10.7566/JPSJ.92.035001}{\emph{J. Phys. Soc. Jap.} {\bfseries 92} (2023) 035001} [\href{https://arxiv.org/abs/2305.03449}{{\ttfamily 2305.03449}}].

\bibitem{Boyd:1994tt}
C.G.~Boyd, B.~Grinstein and R.F.~Lebed, \emph{{Constraints on form-factors for exclusive semileptonic heavy to light meson decays}}, \href{https://doi.org/10.1103/PhysRevLett.74.4603}{\emph{Phys. Rev. Lett.} {\bfseries 74} (1995) 4603} [\href{https://arxiv.org/abs/hep-ph/9412324}{{\ttfamily hep-ph/9412324}}].

\bibitem{Boyd:1995cf}
C.G.~Boyd, B.~Grinstein and R.F.~Lebed, \emph{{Model independent extraction of \ensuremath{|V_{cb}|} using dispersion relations}}, \href{https://doi.org/10.1016/0370-2693(95)00480-9}{\emph{Phys. Lett. B} {\bfseries 353} (1995) 306} [\href{https://arxiv.org/abs/hep-ph/9504235}{{\ttfamily hep-ph/9504235}}].

\bibitem{Boyd:1995sq}
C.G.~Boyd, B.~Grinstein and R.F.~Lebed, \emph{{Model independent determinations of \ensuremath{\bar{B}\to D\ell\bar{\nu}}, \ensuremath{D^\star\ell\bar{\nu}} form factors}}, \href{https://doi.org/10.1016/0550-3213(95)00653-2}{\emph{Nucl. Phys. B} {\bfseries 461} (1996) 493} [\href{https://arxiv.org/abs/hep-ph/9508211}{{\ttfamily hep-ph/9508211}}].

\bibitem{Boyd:1997kz}
C.G.~Boyd, B.~Grinstein and R.F.~Lebed, \emph{{Precision corrections to dispersive bounds on form-factors}}, \href{https://doi.org/10.1103/PhysRevD.56.6895}{\emph{Phys. Rev. D} {\bfseries 56} (1997) 6895} [\href{https://arxiv.org/abs/hep-ph/9705252}{{\ttfamily hep-ph/9705252}}].

\bibitem{Grinstein:2017nlq}
B.~Grinstein and A.~Kobach, \emph{{Model-Independent Extraction of $|V_{cb}|$ from $\bar{B}\rightarrow D^* \ell \overline{\nu}$}}, \href{https://doi.org/10.1016/j.physletb.2017.05.078}{\emph{Phys. Lett. B} {\bfseries 771} (2017) 359} [\href{https://arxiv.org/abs/1703.08170}{{\ttfamily 1703.08170}}].

\bibitem{Caprini:1997mu}
I.~Caprini, L.~Lellouch and M.~Neubert, \emph{{Dispersive bounds on the shape of \ensuremath{\bar{B}\to D^\star \ell\bar{\nu}} form-factors}}, \href{https://doi.org/10.1016/S0550-3213(98)00350-2}{\emph{Nucl. Phys. B} {\bfseries 530} (1998) 153} [\href{https://arxiv.org/abs/hep-ph/9712417}{{\ttfamily hep-ph/9712417}}].

\bibitem{Huang:2022qsb}
Z.~Huang, E.~Gull and L.~Lin, \emph{{Robust analytic continuation of Green's functions via projection, pole estimation, and semidefinite relaxation}}, \href{https://doi.org/10.1103/PhysRevB.107.075151}{\emph{Phys. Rev. B} {\bfseries 107} (2023) 075151} [\href{https://arxiv.org/abs/2210.04187}{{\ttfamily 2210.04187}}].

\bibitem{Yu:2024ncm}
Y.~Yu, A.F.~Kemper, C.~Yang and E.~Gull, \emph{{Denoising of imaginary time response functions with Hankel projections}}, \href{https://doi.org/10.1103/PhysRevResearch.6.L032042}{\emph{Phys. Rev. Res.} {\bfseries 6} (2024) L032042} [\href{https://arxiv.org/abs/2403.12349}{{\ttfamily 2403.12349}}].

\bibitem{Huang:2023gpb}
L.~Huang and S.~Liang, \emph{{Reconstructing lattice QCD spectral functions with stochastic pole expansion and Nevanlinna analytic continuation}}, \href{https://doi.org/10.1103/PhysRevD.109.054508}{\emph{Phys. Rev. D} {\bfseries 109} (2024) 054508} [\href{https://arxiv.org/abs/2309.11114}{{\ttfamily 2309.11114}}].

\bibitem{Bernecker:2011gh}
D.~Bernecker and H.B.~Meyer, \emph{{Vector Correlators in Lattice QCD: Methods and applications}}, \href{https://doi.org/10.1140/epja/i2011-11148-6}{\emph{Eur. Phys. J. A} {\bfseries 47} (2011) 148} [\href{https://arxiv.org/abs/1107.4388}{{\ttfamily 1107.4388}}].

\bibitem{Patella:2024cto}
A.~Patella and N.~Tantalo, \emph{{Scattering Amplitudes from Euclidean Correlators: Haag-Ruelle theory and approximation formulae}},  \href{https://arxiv.org/abs/2407.02069}{{\ttfamily 2407.02069}}.

\bibitem{Bruno:2024fqc}
M.~Bruno, L.~Giusti and M.~Saccardi, \emph{{Spectral densities from Euclidean lattice correlators via the Mellin transform}},  \href{https://arxiv.org/abs/2407.04141}{{\ttfamily 2407.04141}}.

\bibitem{Wagman:2024rid}
M.L.~Wagman, \emph{{Lanczos, the transfer matrix, and the signal-to-noise problem}},  \href{https://arxiv.org/abs/2406.20009}{{\ttfamily 2406.20009}}.

\bibitem{Hackett:2024nbe}
D.C.~Hackett and M.L.~Wagman, \emph{{Block Lanczos for lattice QCD spectroscopy and matrix elements}},  \href{https://arxiv.org/abs/2412.04444}{{\ttfamily 2412.04444}}.

\bibitem{Hackett:2024xnx}
D.C.~Hackett and M.L.~Wagman, \emph{{Lanczos for lattice QCD matrix elements}},  \href{https://arxiv.org/abs/2407.21777}{{\ttfamily 2407.21777}}.

\end{thebibliography}\endgroup

\end{document}